\documentclass{article}
\usepackage{spconf,amsmath,graphicx}
\usepackage{tabularx}
\usepackage{adjustbox}
\usepackage{multirow}
\usepackage{enumitem}
\usepackage{float}
\usepackage[colorlinks=true, allcolors=blue]{hyperref}
\usepackage{amsfonts}

\title{EMC\textsuperscript{2}-Net: JOINT EQUALIZATION AND MODULATION CLASSIFICATION\\
BASED ON CONSTELLATION NETWORK}
\name{Hyun Ryu and Junil Choi}
\address{School of Electrical Engineering\\
Korea Advanced Institute of Science and Technology\\
\{ryuhyun1905, junil\}@kaist.ac.kr}
\begin{document}
%
\maketitle
\begin{abstract}
Modulation classification (MC) is the first step performed at the receiver side unless the modulation type is explicitly indicated by the transmitter. Machine learning techniques have been widely used for MC recently.
In this paper, we propose a novel MC technique dubbed as Joint \textbf{E}qualization and \textbf{M}odulation \textbf{C}lassification based on \textbf{C}onstellation Network (\texttt{EMC\textsuperscript{2}-Net}).
Unlike prior works that considered the constellation points as an image, the proposed \texttt{EMC\textsuperscript{2}-Net} directly uses a set of 2D constellation points to perform MC. In order to obtain clear and concrete constellation despite multipath fading channels, the proposed \texttt{EMC\textsuperscript{2}-Net} consists of equalizer and classifier having separate and explainable roles via novel three-phase training and noise-curriculum pretraining. Numerical results with linear modulation types under different channel models show that the proposed \texttt{EMC\textsuperscript{2}-Net} achieves the performance of state-of-the-art MC techniques with significantly less complexity.
\end{abstract}
\section{Introduction}
\label{sec:intro}
In the case of wireless communication in general, the transmitter side gives information of modulation type (MT) to the receiver side. It also allows the receiver side to equalize a signal by acquiring channel information through pilot signals. There are some systems, however, that MT is not given to the receiver side, for instance, signal confirmation, interference identification, and surveillance in civilian and military applications \cite{yu2006automatic, dobre2007survey}. It is important to identify precise MT on the receiver side for these systems, which is a challenging problem to solve.\\
\indent \textit{Why is MC challenging?} There are two scenarios when MT is hard to recognize.
First is the case when the level of modulation gets higher (e.g., 256-QAM). Higher-order modulation means a large number of clusters exist, and the distance between each symbol gets closer in normalized I/Q coordinates. Thus, to achieve the same bit error rate, higher-order modulation needs a higher signal-to-noise ratio (SNR) value than that of the lower one.
Second is the case when signal quality is poor due to (but not limited to) low SNR and severe fading.
At low SNR, each cluster of symbols overlaps with one another, which makes it difficult to recognize the number of clusters and overall distribution of the constellation.
With severe fading, proper equalization should be preceded, making the constellation 2D Gaussian Mixture (GM) \cite{laot2005closed}. Without any pilot signal, however, blind channel equalization is the only option.\\
\indent \textit{Conventional MC} consists of two separate steps, blind channel equalization followed by MC.
The constant modulus algorithm (CMA) \cite{godard1980self} is one of the most commonly used blind channel equalization techniques. CMA equalizes a signal only with a given MT, but it has two significant drawbacks. One, it is a time-consuming, computationally expensive, iterative algorithm. Two, the output quality of CMA varies much depending on the choice of hyperparameters.
Once a signal is equalized, there are two methodologies for MC, likelihood-based (LB) and feature-based (FB) methods.
For LB, the constellation fits one of the Gaussian probability density functions of target MTs as in maximum likelihood (ML) \cite{wei2000maximum}.
For FB, higher order cumulants (HOCs) of a signal are fed to decision algorithms such as multilayer perceptron (MLP) \cite{xie2019deep}. However, LB is still time-consuming and requires additional measurement of SNR value, while FB shows degraded classification accuracy on higher-order modulation.\\
\indent \textit{Deep learning-based MC} resolves the time-consuming issue yet achieves promising performance.
Most prior works focused on how to design a convolutional neural network (CNN) with 1D filters \cite{o2016convolutional, ramjee2019fast, huynh2020mcnet, huynh2020chain, tunze2020sparsely} or recurrent neural network (RNN) \cite{zhang2020automatic, lin2020hybrid} where sequential signals are input. The problem is that these did not exploit any domain knowledge which questions how neural network (NN) did their job.
Some few works took constellation images as input of CNN with 2D filters \cite{wang2019data}. It causes, however, significant information loss because image resolution limits the expressivity of constellation points.\\
\indent In this paper, we propose Joint \textbf{E}qualization and \textbf{M}odul-ation \textbf{C}lassification based on \textbf{C}onstellation Network (\texttt{EMC\textsuperscript{2}-\\Net}), a new algorithm for MC under the channel with multipath fading followed by additive white Gaussian noise (AWGN).
The source code used in the paper is available at \url{https://github.com/Hyun-Ryu/emc2net}.
The main contributions of our work are:
\begin{enumerate}
    \item Understand constellation as a set of 2D points (in I/Q coordinates) rather than as an image as in prior works, having no information loss.
    \item Train equalizer and classifier jointly under the supervision of MT label only. Two NNs perform completely separate and explainable roles by novel \textit{three-phase training} and \textit{noise-curriculum pretraining}.
    \item \texttt{EMC\textsuperscript{2}-Net} shows state-of-the-art performance on the classification of linear MTs compared to baselines and ablations under three different channels with much less complexity.
\end{enumerate}

\section{Methods}
\label{sec:methods}

\subsection{System Model}
\label{ssec:sysmod}
The overall system consists of the transmitter, channel, and receiver.
On the transmitter side, a sequence of random bits is generated, modulated by one of the target MTs, normalized to have unit power, upsampled by $N$, and pulse-shaped. The length $d$ transmit signal is denoted as $\textbf{x} \in \mathbb{R}^{d}$.
Multipath fading channel is either Rician or Rayleigh which their finite impulse responses are denoted as $\textbf{h} \in \mathbb{R}^{L_{h}}$ with length $L_{h}$. Channel output $\textbf{y} \in \mathbb{R}^{d+L_{h}-1}$ is described as $\textbf{y}=\textbf{h} \ast \textbf{x}+\textbf{n}$ where $\ast$ is convolution and $\textbf{n} \in \mathbb{R}^{d+L_{h}-1}$ is AWGN.
On the receiver side, transients from the beginning of $\textbf{y}$ are removed, trimmed to length $L$, and normalized to unit power, which is denoted as $\textbf{y}_{L} \in \mathbb{R}^{L}$.
This is the way we generate synthetic datasets that are used to train NN.
Moreover, we consider dataset with fading $\textbf{h}$ replaced by random phase offset (PO) which is called AWGN+PO dataset, and the reason for this additional dataset will become clear in Section \ref{sssec:3ptrain}.

\begin{figure}[t]
\centering
\centerline{\includegraphics[width=8.5cm]{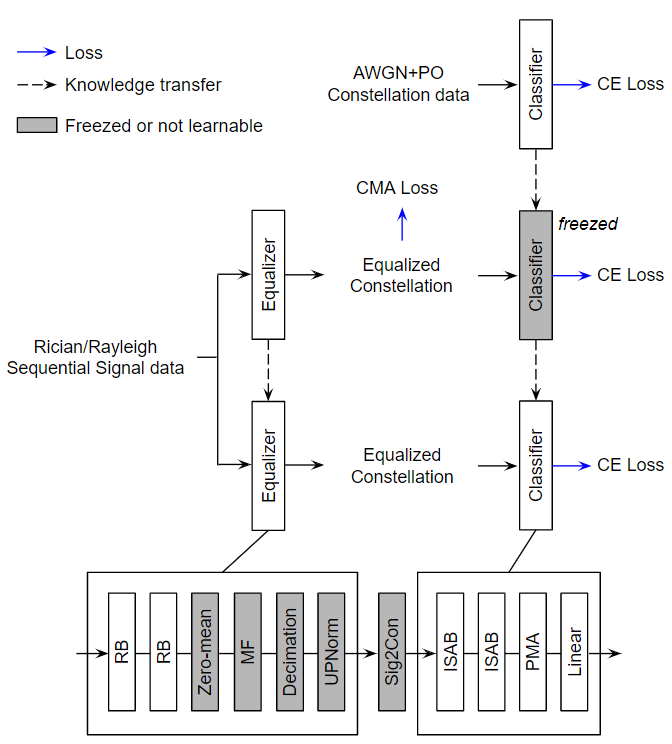}}
\caption{Overall architecture of \texttt{EMC\textsuperscript{2}-Net} and three-phase training strategy.}
\label{fig:emc2net}
\end{figure}

\subsection{\texttt{EMC\textsuperscript{2}-Net} Architecture}
\label{ssec:arch}
\texttt{EMC\textsuperscript{2}-Net} is composed of equalizer and classifier as shown in Fig. \ref{fig:emc2net}.
The equalizer begins with two cascaded residual blocks (RBs), each consisting of Conv1D layers with skip connection. Following the similar logic of prior work on blind channel equalization in \cite{caciularu2018blind}, the number of feature maps in each layer is preserved to 2 to indicate the I/Q channels.
Its output is zero-meaned in order to eliminate the DC offset introduced by RBs.
Matched filter (MF) is designed to match the pulse shaping filter on the transmitter side.
The decimation factor is simply the same as the upsampling factor $N$, i.e., the ratio between the sampling rate and the symbol rate. Note that the MF and decimation factor are known prior, i.e., we do not have to train these components. UPNorm gives unit-power normalized output.
The equalizer output, a sequential signal, is converted into constellation (Sig2Con).\\
\indent The classifier starts with two induced set attention blocks (ISABs) \cite{lee2019set} which are feature extractors of a set of constellation points. It is followed by a pooling by multi-head attention (PMA) \cite{lee2019set} which aggregates learned features into low-dimensional space. The last linear layer eventually gives the output probabilities of each MT.
The intuition behind how the classifier recognizes constellation is that points in the same cluster show high interaction, whereas points in different clusters show low interaction. Actual interaction occurs in high dimensional space via the attention mechanism \cite{vaswani2017attention} which, in a physical sense, selects the nearest points around each point in the constellation.

\subsection{\texttt{EMC\textsuperscript{2}-Net} Training}
\label{ssec:train}

\subsubsection{Three-phase training strategy}
\label{sssec:3ptrain}
We propose a novel three-phase training strategy, illustrated in Fig. \ref{fig:emc2net}, in order to separate the roles of the equalizer and classifier.\\
\indent Phase 1 is \textit{classifier pretraining}. Presuming the ideal equalizer, the classifier input is the perfect GM constellation. PO which inevitably occurs in communication, however, cannot be recovered by equalization such as CMA, which is why the AWGN+PO dataset is considered in Phase 1.
In order to train the classifier in a robust manner, inspired by \cite{bengio2009curriculum}, we apply noise-curriculum based on SNR values in which constellation points of the highest SNR are given first then those of low SNR are gradually added to the dataset as training goes by.\\
\indent Phase 2 is \textit{equalizer training}. The equalizer corrects fading, which is independent to classify the GM constellation. It leads us to freeze the classifier while training the equalizer on fading datasets, either Rayleigh or Rician.\\
\indent Phase 3 is \textit{equalizer-classifier fine-tuning}. The equalizer output is GM if the equalizer works ideally, but empirically it is barely achieved. Freezing the classifier in Phase 2 may limit overall performance since its input is no more GM as in Phase 1. Therefore, the classifier needs to adapt to the equalizer output distribution.
There is one concern that if the classifier finds an undesirable local minimum, then the equalizer output can be far apart from the GM of target MTs. We resolve this issue by keeping the learning rate of the classifier smaller than that of the equalizer to prevent severe forgetting of the classifier.

\subsubsection{Loss functions}
\label{sssec:loss}
Phase 1 and 3 are simply guided by cross-entropy (CE) loss.
What we focus on is Phase 2. Denote the equalizer as $f(\cdot)$, its output $\hat{\textbf{z}}$ is described as $\hat{\textbf{z}}=f(\textbf{y}_{L})$. Let the classifier be $g(\cdot)$, its output $\hat{\textbf{p}}$ is described as $\hat{\textbf{p}}=g(\hat{\textbf{z}})$.
By denoting $a_{i}$ as the $i$-th element of vector $\mathbf{a}$, the total loss is then defined as follows:
\begin{align*}
L_{\mathrm{ce}}=-\Sigma_{i=1}^{C} p_i\cdot \log(\hat{p_i}), \quad L_{\mathrm{cma}}&=(|\hat{\textbf{z}}|^2-R_2)^2\\
L_{\mathrm{tot}}=L_{\mathrm{ce}}+\lambda\cdot L_{\mathrm{cma}}
\end{align*}
where $\textbf{p}$ is the MT label in an one-hot vector, $R_{2}=E\{|\textbf{z}|^4\}/\\E\{|\textbf{z}|^2\}$, $\textbf{z}$ is the normalized ground truth constellation of $\textbf{p}$, $C$ is the number of classes, i.e., the number of possible MTs, and $\lambda$ is the balancing hyperparameter. CMA loss \cite{wang2009generalized} is additionally included to enforce the Gaussianity of the equalized constellation. It corrects non-grid-like clusters in the constellation to be grid-like which makes all the clusters distinctive. It is excluded in Phase 3 to give better classification accuracy.

\section{Experiments}
\label{sec:exp}

\subsection{Synthetic Datasets}
\label{ssec:data}
We consider 8 target linear MTs: BPSK, QPSK, 8PSK, 16QAM, 32QAM, 64QAM, 128QAM, and 256QAM where 32QAM and 128QAM are cross-shaped and the other three QAMs are square-shaped. Upsampling factor $N$ is 8, and a root-raised cosine filter with a roll-off factor of 0.35 which spans 4 symbols is used as the pulse shaping filter. The transmitter output length $d$ is 16,384.
Channel specifications are sampling rate 200 kHz, path delays [0, 9, 17] $\mu s$, average path gains [0, -2, -10] dB, maximum Doppler shift 4 Hz, and K-factor 4 for Rician, 0 for Rayleigh fading. Note that each path consists of $L_{h} = 18$ taps with the same specifications.
SNR is kept at 30 dB unless stated otherwise.
Trimmed signal frame length $L$ is 8,192. Each frame contains $L/N = 1,024$ symbols which is a sufficient number for each cluster in the constellation even in higher-order modulations, for instance, an average of 4 symbols per cluster on 256QAM. The total number of frames is 16K, i.e., 8 MTs with 2K frames each.
Furthermore, the AWGN+PO dataset shares the same parameters while a random PO is introduced on each frame instead of fading channel and SNR varies from 10 to 28 dB with a 2 dB interval rather than a fixed value.
Data is generated via MATLAB and it is divided into train, validation, and test sets by 80\%, 10\%, and 10\%, respectively.

\subsection{Baselines and Ablations}
\label{ssec:base}
There are four types of baselines: Conventional (ML, HOC-MLP), CNN-based (CNN1D, ResNet1D, MCNet, ChainNet, SCGNet), RNN-based (CNN-LSTM, HybridNet), and Image-based (CNN2D, ResNet2D) methods.\\
\indent For ML \cite{wei2000maximum}, given SNR, each data is compared to every constellation of target MTs with varying POs, which aims to find an MT with maximum probability. Since it is not directly applicable to fading channels, CMA is preceded by ML on the fading datasets.
For HOC-MLP \cite{xie2019deep}, total 9 cumulants are fed to 3-layer MLP with ReLU activations.
For CNN/RNN-based methods \cite{o2016convolutional, ramjee2019fast, huynh2020mcnet, huynh2020chain, tunze2020sparsely, zhang2020automatic, lin2020hybrid}, NNs are modified to enable 8,192-length signals on our problem setting. Specifically, RNN-based NNs consist of CNN followed by RNN since pure RNNs cannot afford such long signals. Those reduce the temporal size by pooling or large stride in the former CNN.
For Image-based methods \cite{wang2019data, he2016deep}, the input size is fixed by 128$\times$128 with grayscale.\\
\indent Ablation studies on two perspectives, i.e., training procedure and equalization method, are also performed.
Each one of the three phases is excluded during training in order to validate whether all three phases are necessary.
We also consider \texttt{EMC\textsuperscript{2}-Net} by replacing the equalizer with CMA to empirically show the learnable equalizer outperforms CMA.

\subsection{Training Details}
\label{ssec:det}
The total loss with the balancing hyperparameter $\lambda = 10^{-2}$ is minimized by the Adam optimizer. It has a learning rate of $10^{-3}$ which updates the parameters of the entire network at every phase, except that the learning rate of the classifier at Phase 3 is $2.5 \times 10^{-4}$.
The batch size is 64 and the network is trained by 500 epochs.
The noise-curriculum in Phase 1 is as follows: starting from 28 dB, every time 10 epochs passed, 2 dB lower SNR data is continuously added to the training set until the 90\textsuperscript{th} epoch, and all data is given after.
The Conv1D layer of the equalizer has a kernel size of 65. On the classifier, ISABs have 128 hidden dimensions, 4 attention heads, and 64 inducing points, while PMA has a single seed vector. Both dropouts right before PMA and linear layer have a 0.5 probability.
Pytorch \cite{paszke2019pytorch} was used to implement the whole network, and experiments were performed on a single NVIDIA RTX 3090 GPU.

\begin{table}[t]
\centering
\caption{Classification accuracy and computational complexity of models under three datasets. Top-1s are colored blue, and Top-3s are marked bold for each dataset. The asterisk $^\ast$ on baselines indicates constellation is input.\\}\label{tab1}
\begin{adjustbox}{width=\columnwidth}
\begin{tabular}{lccccc}
\hline
\multirow{2}{*}{Method} & Params & Runtime & Rician & Rayleigh & AWGN\\
 & (K) & (ms) & (\%) & (\%) & +PO(\%)\\
\hline
ML$^\ast$ \cite{wei2000maximum}  &  N/A & N/A & 68.59 &  62.78 &  \color{blue}\textbf{96.32}\\
HOC-MLP \cite{xie2019deep}  & 20 & N/A &  65.44 &   63.59 &  78.04\\
\hline
CNN1D \cite{o2016convolutional}  & 167,898 & 0.030 &   53.09 &   49.97 &   60.26\\
ResNet1D \cite{ramjee2019fast}  & 1,175 & 0.209 &   83.72 &   82.91 &   92.17\\
MCNet \cite{huynh2020mcnet}  & 207 & 0.317 &   \textbf{88.50} &   \textbf{86.44} &   92.38\\
ChainNet \cite{huynh2020chain}  & 331 & 0.176 &   82.84 &   84.09 &   84.86\\
SCGNet \cite{tunze2020sparsely}  & 441 & 1.704 &   83.00 &   81.34 &   91.42\\
\hline
CNN-LSTM \cite{zhang2020automatic}  & 1,150 & 0.310 &   84.62 &   81.22 &   89.86\\
HybridNet \cite{lin2020hybrid}  & 983 & 0.396 &   85.31 &   83.94 &  \textbf{94.22}\\
\hline
CNN2D$^\ast$ \cite{wang2019data}  & 8,044 & 0.042 &   68.97 &   69.12 &   92.31\\
ResNet2D$^\ast$ \cite{he2016deep}  & 602 & 0.252 &   66.78 &   67.09  &   92.74\\
\hline
\texttt{EMC\textsuperscript{2}-Net}  & 300 & 0.120 & \color{blue}\textbf{89.16} & \color{blue}\textbf{86.53} & \textbf{93.35}\\
\texttt{EMC\textsuperscript{2}-Net} w/o P1  & 300 & 0.120 & 85.75 &   83.19 &   N/A  \\
\texttt{EMC\textsuperscript{2}-Net} w/o P2  & 300 & 0.120 & \textbf{87.66} & \textbf{86.28} &   N/A  \\
\texttt{EMC\textsuperscript{2}-Net} w/o P3  & 300 & 0.120 &   80.81 &   72.38 &   N/A  \\
\texttt{EMC\textsuperscript{2}-Net} w/ CMA  & 299 & N/A &   86.53 &   81.53 &   N/A  \\
\hline
\end{tabular}
\end{adjustbox}
\label{table1}
\end{table}

\begin{figure}[t]
\centering
\centerline{\includegraphics[width=\columnwidth]{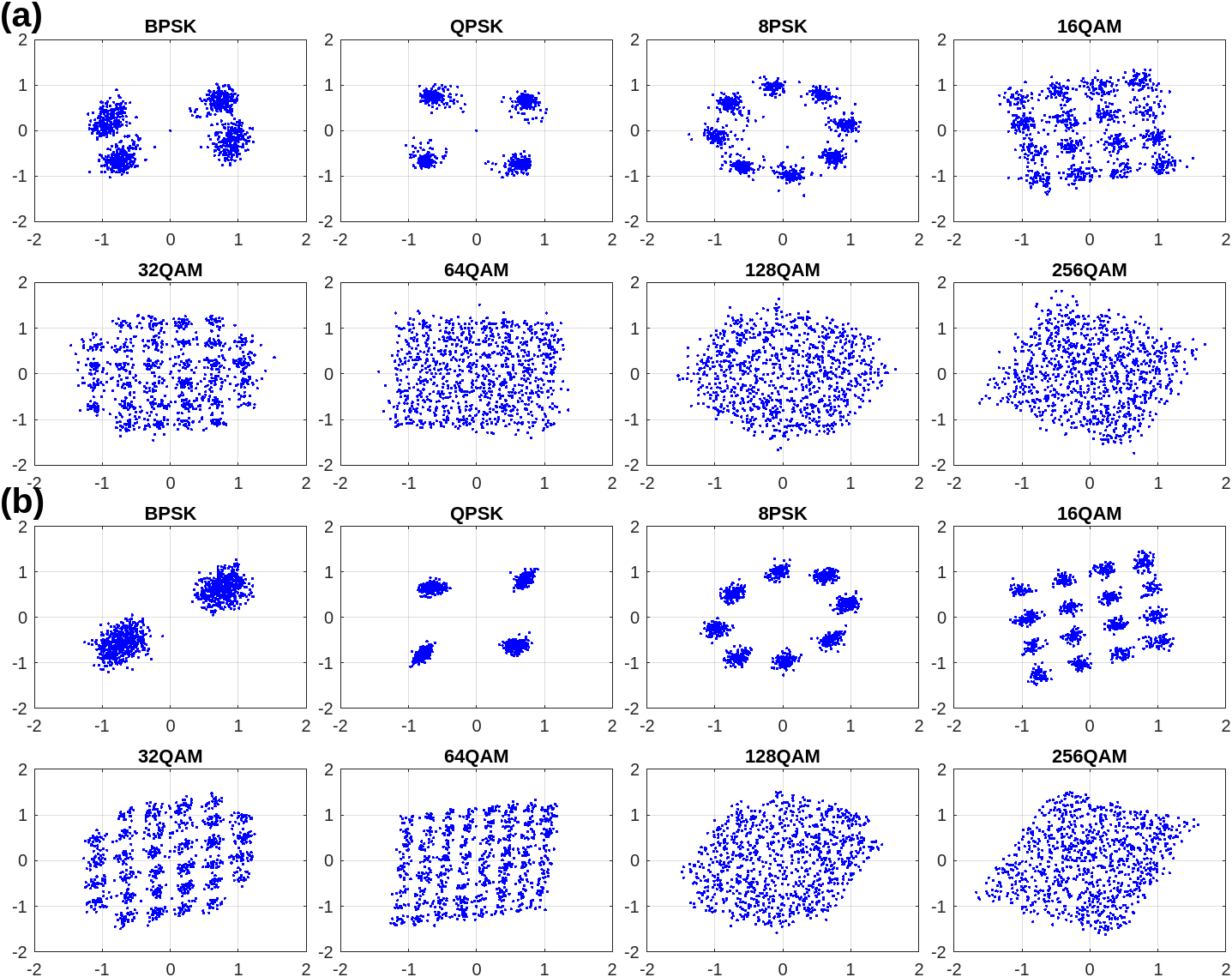}}
\caption{The equalizer output of (a) CMA and (b) \texttt{EMC\textsuperscript{2}-Net} of corresponding signals for each linear modulation where top row shows BPSK, QPSK, 8PSK, 16QAM and bottom one shows 32QAM, 64QAM, 128QAM, 256QAM.}
\label{fig:eq_ours_cma}
\end{figure}

\section{Results}
\label{sec:res}
As summarized in Table \ref{table1}, the proposed \texttt{EMC\textsuperscript{2}-Net} outperforms state-of-the-art baselines on the fading datasets which are our main interest. It also gives a notable result on the AWGN+PO dataset, and ablations are omitted since those are not applicable.
\texttt{EMC\textsuperscript{2}-Net} reduces the number of parameters significantly compared with baselines, even with memory-efficient CNNs, ChainNet and SCGNet, which MCNet is the only one that has fewer parameters than \texttt{EMC\textsuperscript{2}-Net}.\\
\indent Baselines in which the constellation is input, ML, CNN2D, and ResNet2D, show prominent or even the best performance on the AWGN+PO dataset whereas their performances are substantially reduced on the fading datasets.
It is because CNN2D and ResNet2D simply convert signals into constellation points without any equalization. ML pre-equalize signals still failed because of unstable convergence of CMA depending on hyperparameters.
Baselines with CNN and RNN which aimed to somehow implicitly learn equalization and MC show promising result but not as much as \texttt{EMC\textsuperscript{2}-Net} do where it learns each process explicitly.
MCNet slightly underperforms \texttt{EMC\textsuperscript{2}-Net} with fewer parameters. \texttt{EMC\textsuperscript{2}-Net}, however, reduces runtime about 3$\times$ than MCNet, because MCNet consists of 52 convolution layers while \texttt{EMC\textsuperscript{2}-Net} consists of 5 convolution layers followed by 5 attention layers.
HybridNet surpasses the proposed method on the AWGN+PO dataset, however, it has over 3$\times$ more parameters and a longer runtime.
The remaining CNN/RNN-based methods consistently underperform \texttt{EMC\textsuperscript{2}-Net} on all datasets.
HOC-MLP shows way inferior results to others due to its innate limitation of explicit feature extraction.\\
\indent Ablations on training strategy show degraded accuracy which demonstrates each phase is essential. In the order of Rician and Rayleigh, excluding Phase 1 diminishes accuracy by 3.41\% and 3.34\%, Phase 2 by 1.50\% and 0.25\%, and Phase 3 by 8.35\% and 14.15\%.
Ablation on the equalization method shows that \texttt{EMC\textsuperscript{2}-Net} gives better accuracy than \texttt{EMC\textsuperscript{2}-Net} with CMA by 2.63\% on Rician and 5.00\% on Rayleigh which means the proposed equalizer works better than CMA.
For a fair comparison, the pretrained classifier is fine-tuned in Phase 3 using CMA output.
As shown in Fig \ref{fig:eq_ours_cma}, CMA completely misequalizes BPSK into QPSK, which may happen since BPSK and QPSK have the same modulus $R_{2} = 1$, in contrast, the proposed equalizer output shows two tangible clusters. In addition, even in higher-order modulations, especially 64QAM, the proposed equalizer gives definite clusters whereas CMA does not.

\section{Conclusions}
\label{sec:conclusions}
We propose \texttt{EMC\textsuperscript{2}-Net}, a new deep learning-based MC technique that unifies the idea of conventional MC.
Deep learning-based methods so far cannot explain how NN classifies signals whereas \texttt{EMC\textsuperscript{2}-Net} can do it by separating the equalizer and classifier, trained jointly.
The proposed \texttt{EMC\textsuperscript{2}-Net} works simply with MT label by novel \textit{three-phase training} and \textit{noise-curriculum pretraining}.
Moreover, \texttt{EMC\textsuperscript{2}-Net} understands constellation as a set of 2D points, which is information itself, instead of its image as in prior works.
It achieves state-of-the-art performance under multipath fading channels yet is computationally efficient.
\bibliographystyle{IEEEbib}
\bibliography{strings,refs}

\end{document}